\documentclass[pdflatex,sn-mathphys-num]{sn-jnl}

\usepackage{graphicx}%
\usepackage{multirow}%
\usepackage{amsmath,amssymb,amsfonts}%
\usepackage{amsthm}%
\usepackage{mathrsfs}%
\usepackage[title]{appendix}%
\usepackage{xcolor}%
\usepackage{textcomp}%
\usepackage{manyfoot}%
\usepackage{booktabs}%
\usepackage{algorithm}%
\usepackage{algorithmicx}%
\usepackage{algpseudocode}%
\usepackage{listings}%

\newcommand{\GeV}{{\ensuremath\rm GeV}}
\newcommand{\ab}{{\ensuremath\rm ab}}
\newcommand{\lb}{\left(}
\newcommand{\rb}{\right)}

\newcommand{\al}{\alpha}
\newcommand{\be}{\beta}

\theoremstyle{thmstyleone}%
\theoremstyle{thmstyletwo}%

\theoremstyle{thmstylethree}%

\begin{document}
\bibliographystyle{hunsrt}
\title{Low mass scalars at $e^+\,e^-$ colliders}


\author*[1]{\fnm{Tania} \sur{Robens}}\email{trobens@irb.hr}



\affil*[1]{\orgdiv{Division of Theoretical Physics}, \orgname{Rudjer Boskovic Institute}, \orgaddress{\street{Bijenicka cesta 54}, \city{Zagreb}, \postcode{10000}, \country{Croatia}}}




\abstract{
  I briefly discuss the search for low mass scalars at Higgs factories as well as available models that render such scalars feasible, where I focus on new developments since the review presented in \cite{Robens:2022zgk} (see also \cite{Robens:2025kaa} for a recent update).\\
RBI-ThPhys-2026-07  
}




\maketitle
\section{Introduction}

In this short review, I try to briefly summarize the state of the art of additional scalar searches at low center-of-mass energy lepton colliders with $\sqrt{s}\,\sim\,240-250\,\GeV$. In particular, I will give an update on the results presented in \cite{Robens:2022zgk,Robens:2025kaa}, commenting on novel searches and model comparisons that have emerged since then. I also refer the reader to \cite{Altmann:2025feg} for a compact summary that served as input to the European Strategy update. 

Low mass scalars exist in a large variety of new physics scenarios (see e.g. \cite{Robens:2025nev} for a recent review), mainly stemming from models with extended scalar sectors, that are not in contradiction to current theoretical and experimental constraints. Although searches for such scenarios exist at the LHC, it is equally interesting to discuss the prospect of finding or excluding such scenarios at upcoming Higgs factories with a focus on center-of-mass energies of 240 to 250 \GeV. Higher center of mass energies, although also of interest, will not be mentioned here and I refer the reader to \cite{Altmann:2025feg} for more details.

\section{Processes at Higgs factories}

\begin{center}
\begin{figure}[htb!]
\begin{center}

\includegraphics[width=0.55\textwidth]{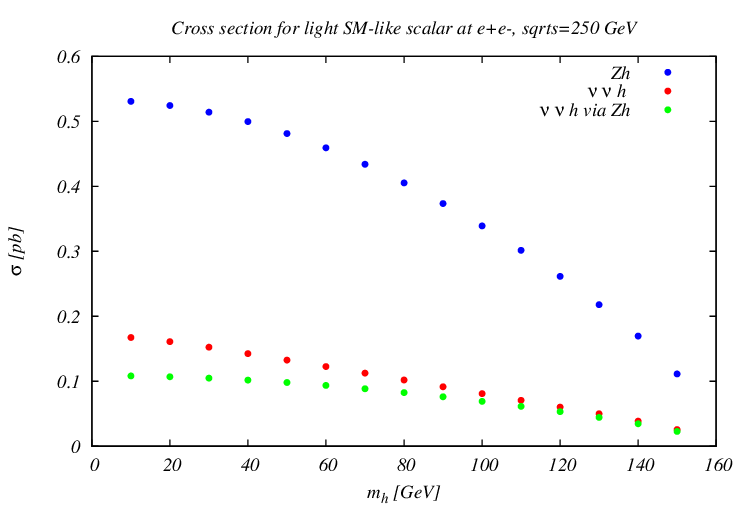}
\caption{\label{fig:prod250} {Leading-order production cross sections for $e^+\,e^-\,\rightarrow\,Z\,h$ {\sl (blue)} and $e^+\,e^-\,\rightarrow\,h\,\nu_\ell\,\bar{\nu}_\ell$ {\sl (red)} production at an $e^+\,e^-$ collider with a com energy of 250 \GeV~using Madgraph5 for an SM-like scalar $h$. We also display the contribution of $Z\,h$ to $\nu_\ell\,\bar{\nu}_\ell\,h$ using a factorized approach for the Z decay, generated via $e^+\,e^-\,\rightarrow\,Z\,h\,\times\,\text{BR}\lb h\,\rightarrow\,\nu_\ell\,\bar{\nu}_\ell \rb$ {\sl (green)}. Taken from \cite{Robens:2024wbw}.}}

\end{center}
\end{figure}
\end{center}

At the center-of-mass (com) energies of Higgs factories, Higgs strahlung is the dominant production mode for single scalar production \cite{Abramowicz:2016zbo}. Leading-order predictions for $Zh$ production at $e^+e^-$ colliders for low mass scalars which are Standard Model (SM)-like, using Madgraph5 \cite{Alwall:2011uj}, are shown in figure \ref{fig:prod250} for a center-of-mass energy of 250 \GeV. The $e^+e^-\,\rightarrow\,h\,\nu_\ell\,\bar{\nu}_\ell$ process contains contributions from both scalar strahlung and VBF type topologies, so we also display the expected rates from the former for this final state using a factorized approach. It can be seen that for higher scalar masses the dominant contribution stems from $Z\,h$ production.

\section{Projections for additional searches and connections to electroweak phase transitions}

This section summarizes the findings already presented in \cite{Robens:2023bzp}. We just list these here as we consider them to be fundamental for the past and current status regarding searches at Higgs factories as well as possible connections to electroweak phase transitions.

In preparation of the European Strategy, a lot of effort was devoted to update several analyses for so-called scalar-strahlung processes at Higgs factories, where a concise summary can e.g. be found in \cite{Altmann:2025feg}. I here briefly
list a couple of sample results and refer the reader to the literature for further details.

In \cite{deBlas:2024bmz}, some target processes were identified that should be reanalysed for the European Strategy input. Although not all of these have been investigated, several new studies exist which we will briefly comment on. These concern mainly scalar-strahlung processes with $b\bar{b}$ and $\tau^+\tau^-$ final states. We also present a comparative study of several of these searches.

\subsection*{$b\,\bar{b}$ final states}

Due to a typically large branching ratio into $b\,\bar{b}$ final states for light scalars in many BSM extensions, this channel is a prime channel to investigate the discovery reach of any $e^+e^-$ collider; in fact, this is a channel that has already been explored in detail at the LEP experiments (see \cite{OPAL:2002ifx,ALEPH:2006tnd} for a concise summary of results.) In general, in these searches there were two different options of treating the decay modes of the additional scalars: one just uses the general recoil against the $Z$ and does not take specific decays of the scalar into account. Another one also uses the final states stemming from the additional scalar\footnote{For a comparison using a LEP recast see e.g. \cite{Drechsel:2018mgd}.}. We here focus on the latter using $b\bar{b}$ final states.

\begin{center}
  \begin{figure}
    \begin{center}
      \includegraphics[width=0.6\textwidth]{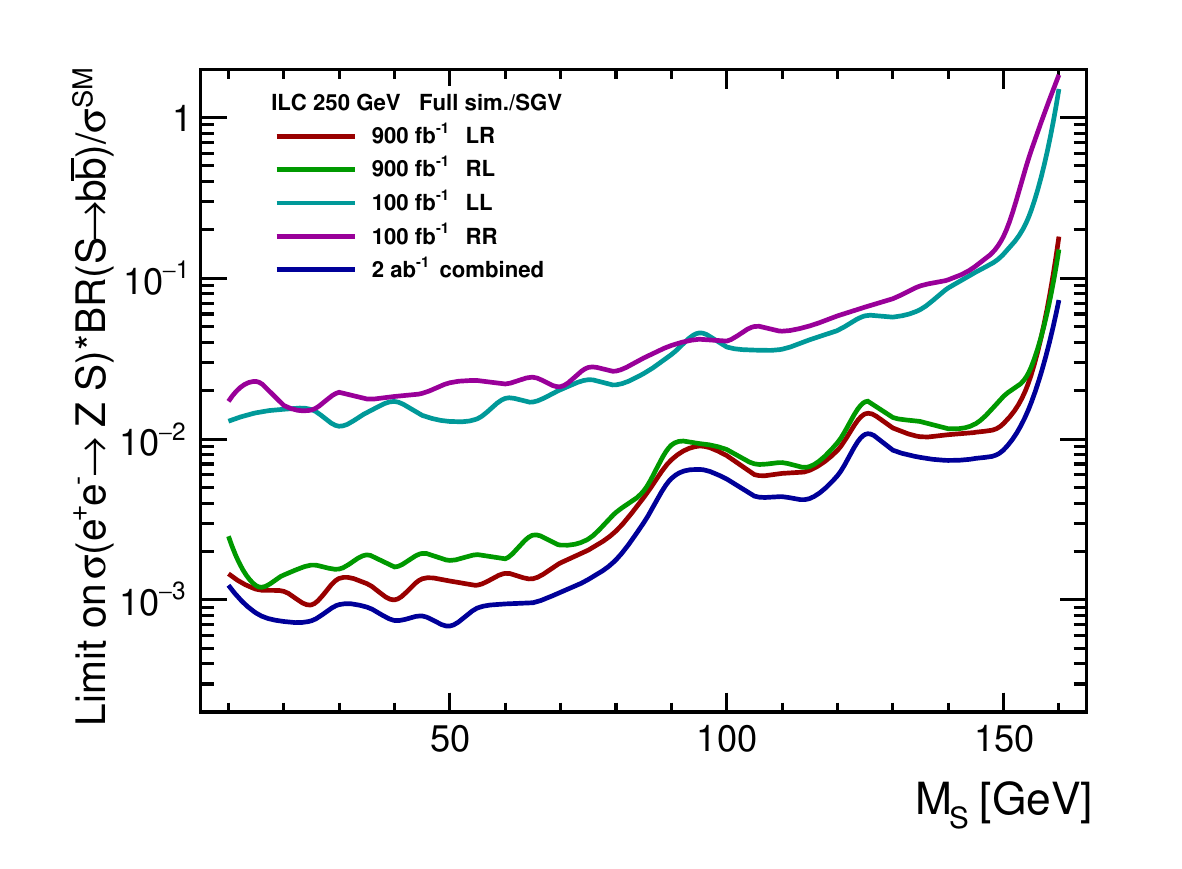}
    \end{center}
    \caption{\label{fig:bb} Taken from \cite{Altmann:2025feg} for $b\bar{b}$ final states, for various initial state polarization combinations at various integrated luminosities as well as a combination at $2\,\ab^{-1}$. Results are given for production times decay, normalized to the SM production cross section at that mass.}
    \end{figure}
  \end{center}

The results of a detailed comparison at a 250 \GeV~ ILC are shown in figure \ref{fig:bb}, including results for several initial state polarization configurations. The best results are achieved for a combination at an integrated luminosity of $2\,\ab^{-1}$. The bound is given on a quantity which can be translated to the squared rescaling for $Z\,S$ production times $S\,\rightarrow\,b\,\bar{b}$ branching ratio in BSM scenarios.

\subsection*{$\tau^+\tau^-$ final states}

\begin{center}
  \begin{figure}
    \begin{center}
      \includegraphics[width=0.49\textwidth]{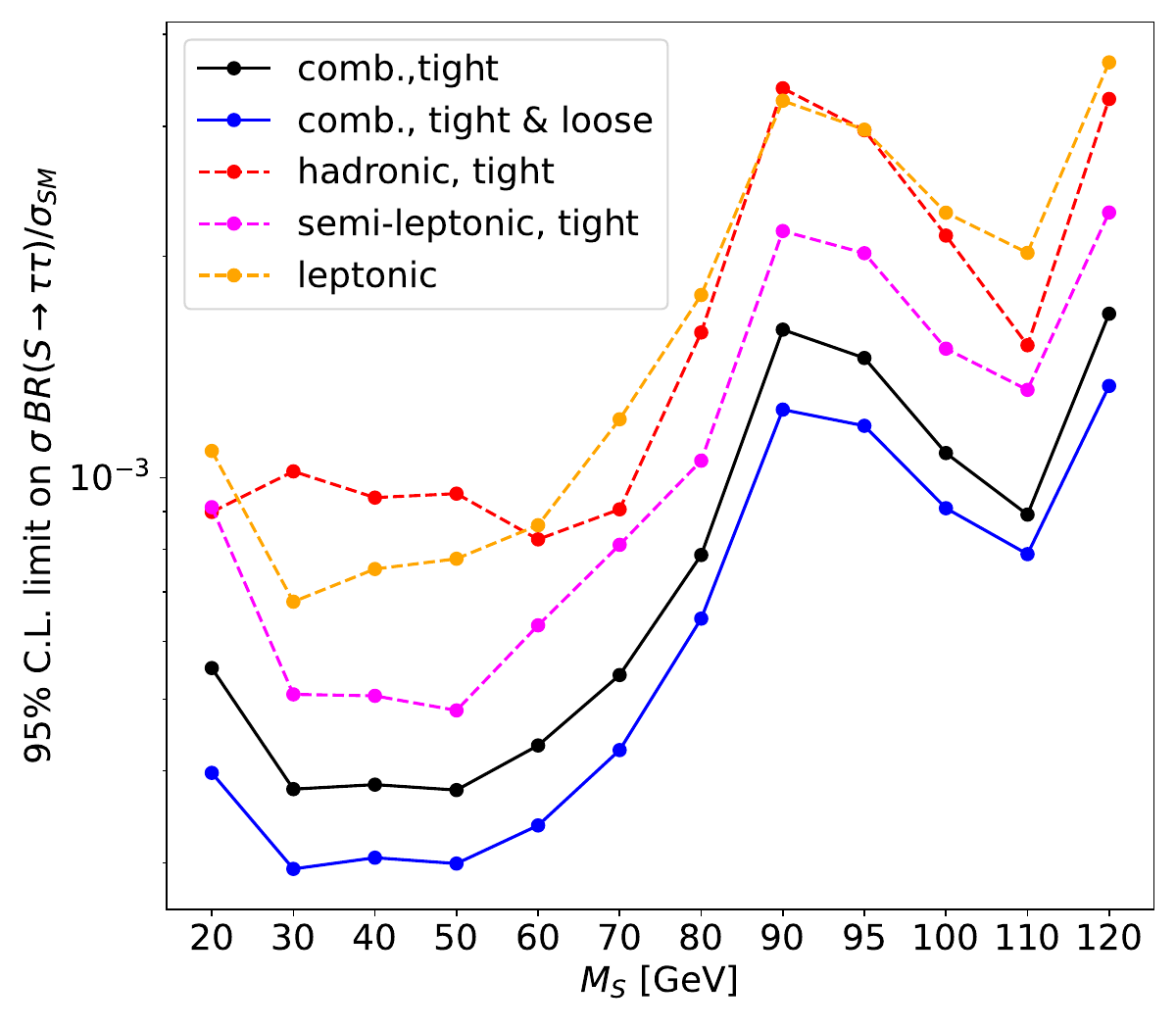}
    \end{center}
    \caption{\label{fig:tautau} Taken from \cite{Altmann:2025feg} for $\tau^+\,\tau^-$ final states, for an ILC at 250 \GeV and an integrated luminosity of $2\,\ab^{-1}$. Result are shown for various final states and selection criteria for the di-tau decay products. See \cite{Altmann:2025feg} for details. }
    \end{figure}
  \end{center}

One of the main targets promoted in \cite{deBlas:2024bmz} were di-tau final states. The results of such a study are shown in figure \ref{fig:tautau}, where various decay and selection modes of the scalar decay products are taken into account. Results are shown again for a center-of-mass energy of 250 \GeV and an ILC setup with an integrated luminosity of $2\,\ab^{-1}$, and given in terms of factorized production times branching ratio rates normalized to the production of an SM-like scalar at that mass. Best results can be obtained in an overall combination of all selection criteria searches.

\subsection*{Comparison}

\begin{center}
  \begin{figure}
    \begin{center}
      \includegraphics[width=0.8\textwidth]{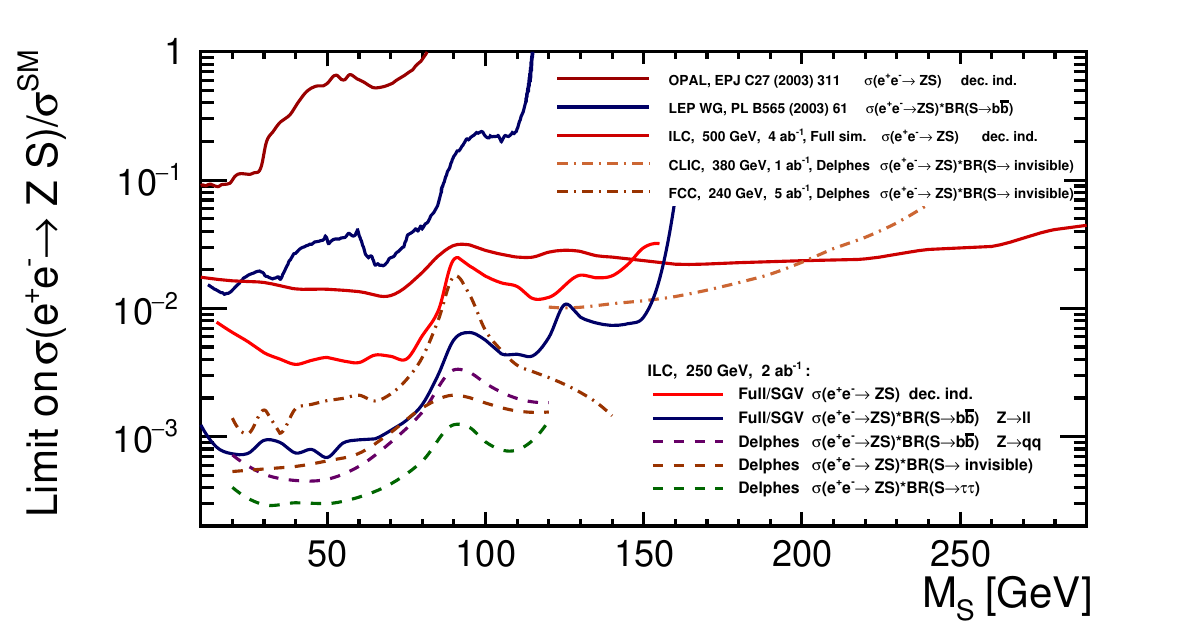}
    \end{center}
    \caption{\label{fig:combi} Taken from \cite{Altmann:2025feg} for several final states, reach of several searches using a variety of decay modes, center-of-mass energies, and detector simulation approaches. Shown are also the experimental bounds achieved by the LEP collaborations \cite{OPAL:2002ifx,ALEPH:2006tnd}.}
    \end{figure}
  \end{center}

Several other final states, in particular also searches with invisible decays of the additional scalar, have been documented in \cite{Altmann:2025feg}. In figure \ref{fig:combi}, we show the comparison of various of such searches and their reach for the normalized rates. Care must be taken in comparing the reaches in this figure, as some results are given for a rescaled production cross section only, while other include a factorized decay for the additional scalar. Higher center-of-mass energies naturally lead to higher mass reaches. Note that this figure also in general includes center of mass energies higher than in the typical Higgs factory setup. For all final states, the 250 \GeV~ ILC studies at 2 $\ab^{-1}$ supersede previous studies for the corresponding mass range. In general the di-tau final states appears to be most sensitive in the studies considered here. Note that the displayed results also differ in terms of the applied detector simulation software. Also displayed are actual search results from LEP.

\begin{center}
\begin{figure}[htb!]
\begin{center}
\includegraphics[width=0.45\textwidth]{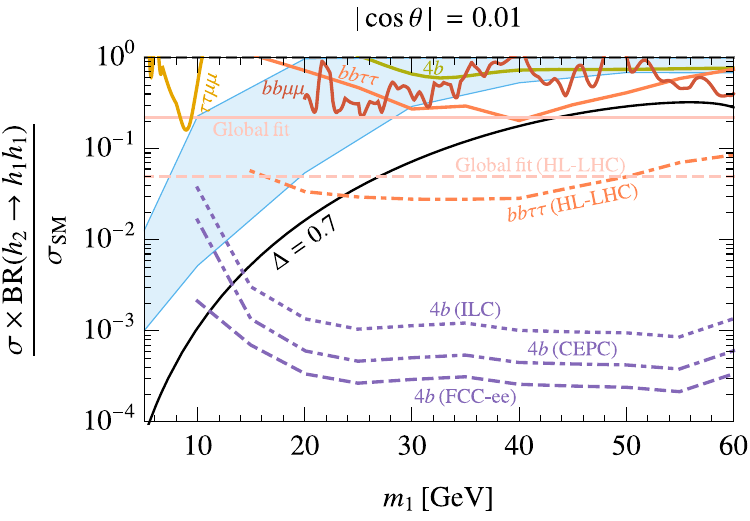}
\includegraphics[width=0.45\textwidth]{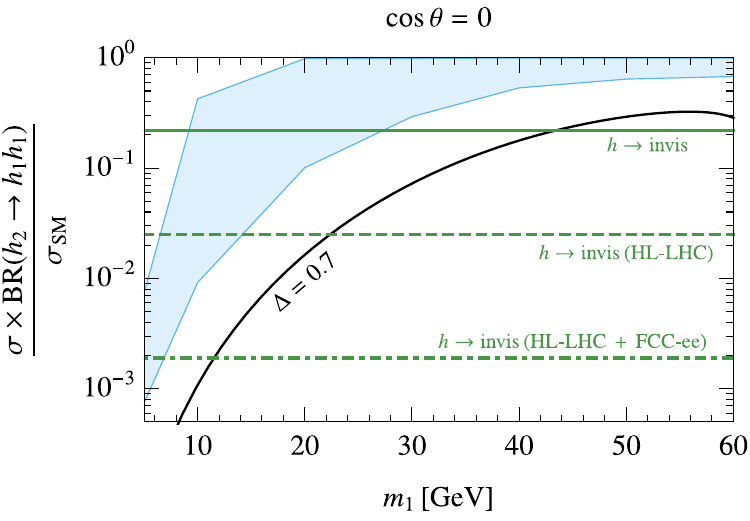}
\end{center}
\caption{\label{fig:ewps} Expected bounds on Higgs production via Higgs strahlung and subsequent decay into two light scalars, in the singlet extension scenario discussed in \cite{Kozaczuk:2019pet,Wang:2022dkz}. For $\cos\theta\,=\,0$ the constraints mainly stem from $h_{125}\,\rightarrow\,\text{invisible}$ searches. The blue shaded area corresponds to the region that allows for a strong first order electroweak phase transition in agreement with the scans performed in \cite{Kozaczuk:2019pet}. $\Delta R$ is a measure of the relative difference of realized singlet vacua to the true vacuum over the difference of the maximal barrier field configuration again with respect to the true vacuum (see text for details). Depending on $m_1$ this scenario can be tested at current or future collider experiments.}
\end{figure}
\end{center}

Another important topic is the connection of models with extended scalar sectors with different scenarios of electroweak phase transitions. In particular, for scenarios where the second scalar is lighter than a SM like candidate, such states can be investigated in Higgs-strahlung. A tell-tale signature in this case is the associated decay of $h\,\rightarrow\,h_i\,h_i$, where $h_i$ denotes a low mass scalar. The clean environment of a lepton collider typically allows to test the respective parameter space into regions that provide a priori relatively low rates. There has a been a lot if recent activity in this field; in figure \ref{fig:ewps}, we here exemplarily show results from \cite{Kozaczuk:2019pet}.

In general, strong first order electroweak phase transitions do not necessarily require a lighter second state. Therefore, regarding the above statement it should be noted that there are also numerous scenarios where the second scalar (or multiple scalars) can be heavier than the 125 \GeV Higgs. However, for production mechanisms at a Higgs factory we would need masses for these particles to be relatively light, $\lesssim\,160\,\GeV$ for on-shell production for example in scalar-strahlung processes. The tell-tale character above therefore refers to the case of light scalars $\lesssim\,62.5\,\GeV$, but other scenarios are of course also possible.

Several collider sensitivity projections are shown, including generic bounds for various discalar decay modes that have originally been derived in \cite{Liu:2016zki}. Show is also the region where strong first order electroweak phase transitions are possible following the analysis in \cite{Kozaczuk:2019pet} {\sl (light blue band)}, as well as a lower bound on the respective normalized cross sections guaranteeing a successful completion of the phase transition, denoted by $\Delta R\,=\,0.7$. It is evident that $e^+e^-$ Higgs factories would be an ideal environment to confirm or rule out such scenarios.

\begin{center}
\begin{figure}
\begin{center}
\includegraphics[scale=0.45]{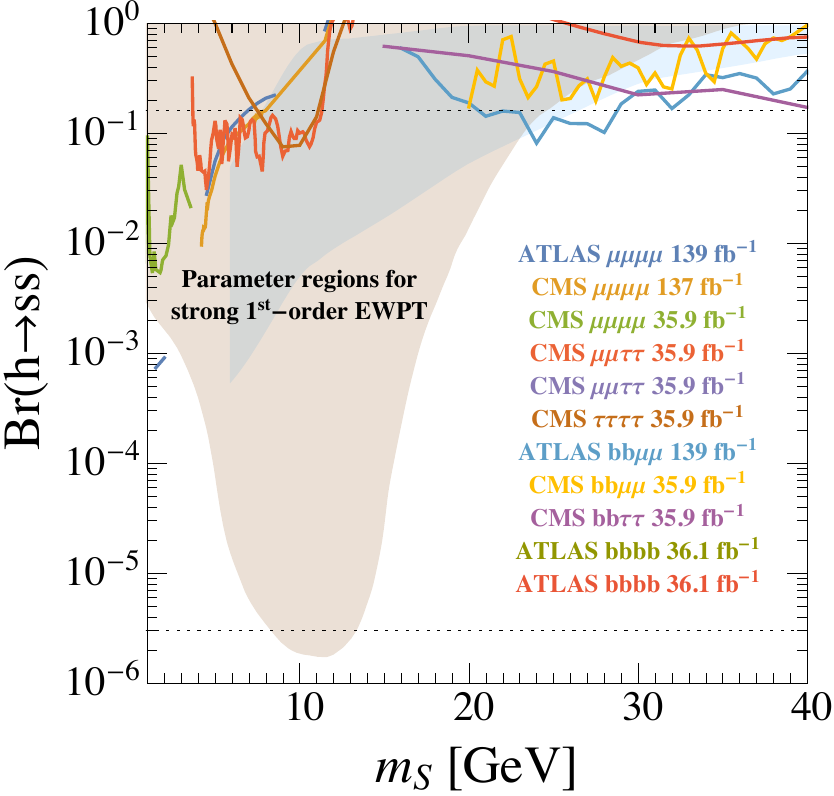}
\includegraphics[scale=0.45]{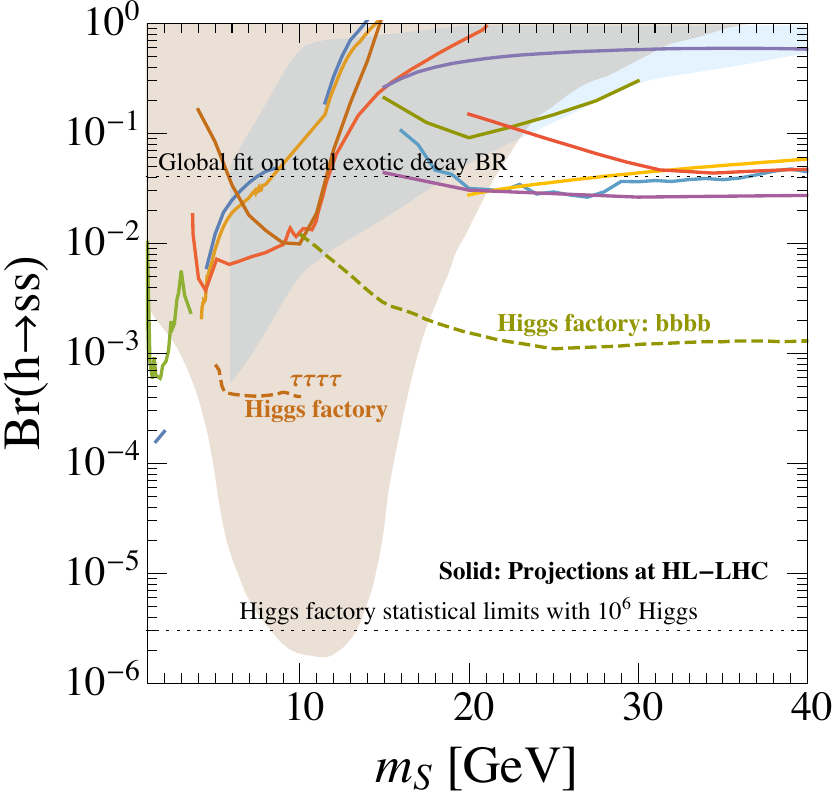}
\caption{\label{fig:snowmass} Regions in the $\lb m_S; \text{Br}\,\lb h\,\rightarrow\,s\,s \rb\rb$ parameter space that allow for a strong first order electroweak phase transition according to the discussion in \cite{Carena:2022yvx}, with a generic {\sl (blue)} singlet extension scenario or a scenario where an additional $\mathbb{Z}_2$ symmetry is broken spontaneously {\sl (beige)}. {\sl Left:} Current constraints and direct searches at the time of the publication. {\sl Right:} Projections for HL-LHC. References for the searches can be found in \cite{Carena:2022yvx}. }
\end{center}
\end{figure}
\end{center}

A similar comparison was done in the context of the Snowmass study, see \cite{Carena:2022yvx}, for various scenarios of symmetries applied to a singlet scenario. Results from by that time current collider bounds as well as Higgs factory projections are displayed in figure \ref{fig:snowmass}.

\section{Parameter space for some sample models}

\subsection*{General discussion}

After briefly discussing new physics signatures, we now turn to models that still allow for such low mass scalars. This is obviously only a brief overview, and more models might exist allong for low mass scalars accessible at Higgs factories; see e.g. \cite{Robens:2022zgk} for more details.

New physics scenarios with additional scalar content typically stem from extending the SM scalar sector by additional singlets/ doublets/ triplets, where the nomenclature decribes the transformation under the SM gauge group. Depending on the specific extension considered, the new physics scenarios contain several CP-even or -odd additional neutral, additional charged, or multi-charged scalars. 

In general, new physics scenarios are subject to a long list of theoretical and experimental constraints, such as conditions on vacuum stability, unitarity, perturbativity of couplings (up to certain scales), as well as experimental findings: current and past collider search results, electroweak precision observables, agreement with flavour data, as well as dark matter findings in case the model contains a dark matter candidate. Another important ingredient is the fact that the new physics scenario has to comply with the properties of the 125 
\GeV~ resonance discovered by the LHC experiments \cite{ATLAS:2022vkf,CMS:2022dwd}.

The main production mode for most models will be scalar strahlung involving the $S Z Z$ vertex, however, we have to mention an important sum rule \cite{Gunion:1990kf} that relates the couplings of additional neutral scalars:

$\sum_i g^2_{h_i V V} \,=\,g^2_{h_\text{SM} V V}$ ,

where the sum runs over all neutral CP-even scalars, $h_\text{SM}$ denotes the SM Higgs, and $VV\,\in\,\left[Z Z; W^+ W^-\right]$. As one of the scalars has to comply with the measurements of the 125 \GeV~ resonance at the LHC experiments, we see that from the signal strength alone we already have strong constraints on the new physics couplings. The above rule is amended in the presence of doubly charged scalars.

\subsection*{Specific models}

\begin{center}
\begin{figure}[htb!]
\begin{center}
\includegraphics[width=0.48\textwidth]{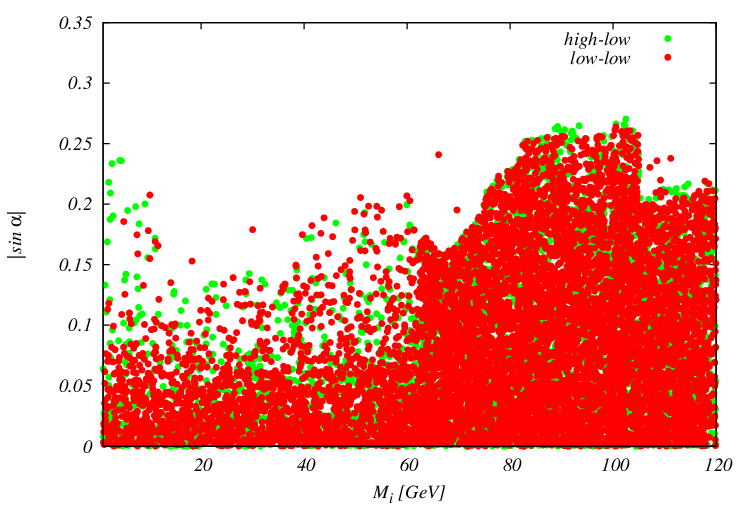}
\includegraphics[width=0.48\textwidth]{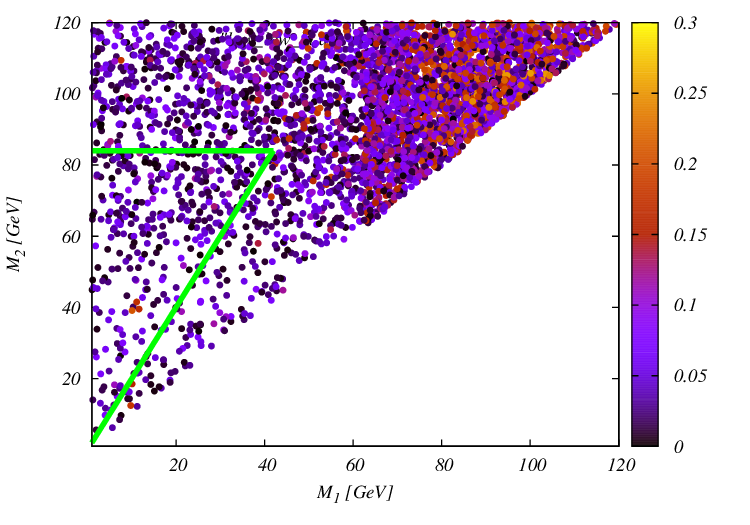}
\caption{\label{fig:trsm} {Available parameter space in the TRSM, with one (high-low) or two (low-low) masses lighter than 125 \GeV. {\sl Left}: light scalar mass and mixing angle, with $\sin\al\,=\,0$ corresponding to complete decoupling. {\sl Right:} available parameter space in the $\lb m_{h_1},\,m_{h_2}\rb$ plane, with color coding denoting the rescaling parameter $\sin\al$ for the lighter scalar $h_1$. Within the green triangle, $h_{125}\,\rightarrow\,h_2 h_1\,\rightarrow\,h_1\,h_1\,h_1$ decays are kinematically allowed. Taken from \cite{Robens:2022zgk}.}}
\end{center}
\end{figure}
\end{center}

The first model we discuss is a model that extends the scalar sector of the SM by two additional fields that transform as singlets under the electroweak gauge group, the Two Real Singlet Model (TRSM) \cite{Robens:2019kga,Robens:2022nnw}. This model contains three CP-even neutral scalars that relate the gauge and mass eigenstates $h_{1,2,3}$ via mixing. One of the scalars necessarily needs to comply with current LHC findings for the SM-like 125 \GeV~ resonance. However, the other two can in principle take any mass values, as long as all current theoretical and experimental constraints are fulfilled.
In figure \ref{fig:trsm}, two cases are shown where either one (high-low) or two (low-low) scalar masses are smaller than $125\,\GeV$. On the y-axis, the respective mixing angle is shown. Decoupling here corresponds to $\sin\al\,=\,0$. The parameter space is basically identical to the one discussed in \cite{Robens:2023bzp}. We use the scan discussed in \cite{Robens:2019kga,Robens:2022nnw}, making use of the ScannerS framework \cite{Coimbra:2013qq,Muhlleitner:2020wwk}. Direct search constraint are implemented via HiggsTools \cite{Bahl:2022igd}.

Another interesting scenario is given by two Higgs doublet models, where the SM scalar sector is augmented by a second doublet. Such models have been studied by many authors, see e.g. \cite{Branco:2011iw} for an overview on the associated phenomenology. We here focus on the available parameter space in 2HDMs of type I, where all leptons couple to one doublet only. In such models, flavour constraints that are severly constraining the parameter space for other Yukawa structures (see e.g. \cite{Misiak:2017bgg} for a discussion in type II models) are leviated.

We here exemplarily show the parameter space for a type I 2HDM, where we fixed all additional scalar masses to the same value $m_H\,=\,m_A\,=\,m_{H^\pm}$ and additionally set $\cos\lb \be-\al\rb$ to a certain value. In one of the two scenarios shown here, we additionally fixed the additional potential parameter $m_{12}^2\,=\,m_H^2\,\sin\be\,\cos\be$.

\begin{center}
\begin{figure}
\begin{center}
\includegraphics[width=0.49\textwidth]{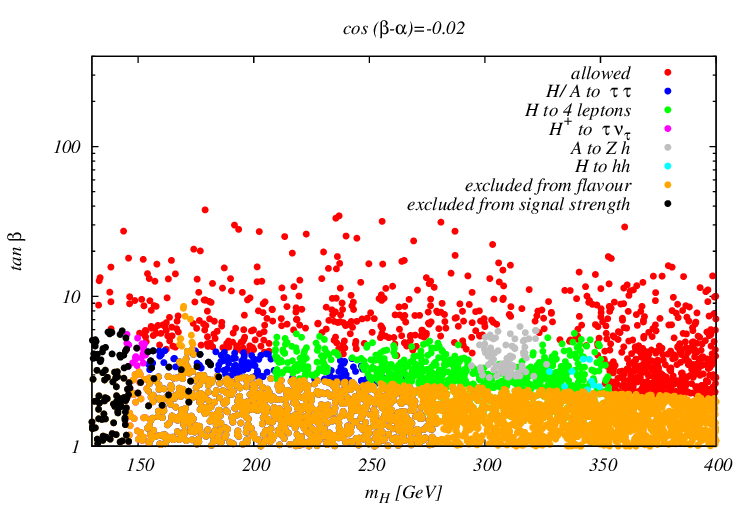}
\includegraphics[width=0.49\textwidth]{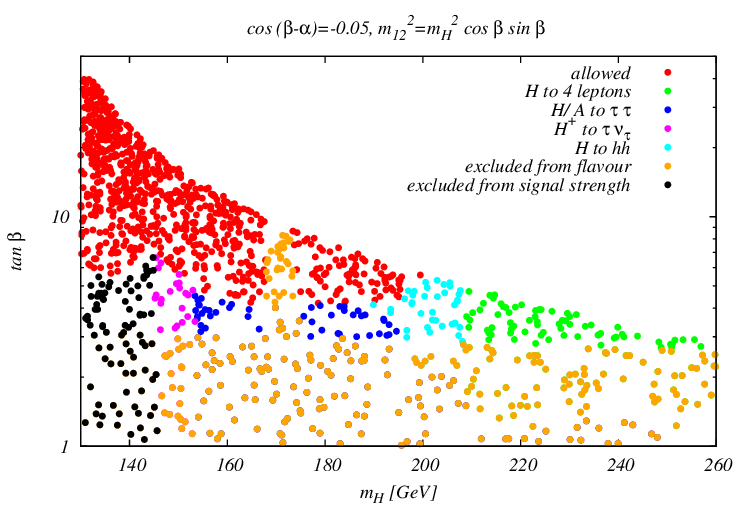}
\caption{\label{fig:2hdm} Allowed regions in the $\lb m_H,\,\tan\be\rb$ plane after various constraints have been taken into account, including experimental searches at the LHC as currently available using thdmTools. {\sl Left:} Here, $\cos\lb \be-\al\rb\,=\,-0.02$, and $m_{12}^2$ is floating freely. {\sl Right:} We here fixed $\cos\lb \be-\al\rb\,=\,-0.05,\,m_{12}^2\,=\,m_H^2\,\sin\be\,\cos\be$. See text for further details.}
\end{center}
\end{figure}
\end{center}

We show the corresponding parameter space in figure \ref{fig:2hdm}, taken from \cite{Robens:2024wbw}. Note that the plot has not been updated to reflect most recent LHC constraints. For a discussion of applied constraints, we refer the reader to \cite{Robens:2024wbw}. The figure has been obtained using thdmTools \cite{Biekotter:2023eil}.

In general, for the two scenarios shown here, the available parameter space is already severely constrained from current searches at the LHC. For the scenario where $\cos\lb \be -\al\rb=-0.02$, current searches cease to be sensitive for masses $\gtrsim\,400\,\GeV$, but we can find allowed parameter points in the whole mass range, where a lower scale for $\tan\be$ is set by flavour constraints. On the other hand, if we additionally fix the potential parameter $m_{12}^2$, we see that there is also an upper limit on $\tan\be$ from the remaining constraints, basically leading to a maximal mass scale of around 200 \GeV~ for the heavy scalars. Note that these bounds only apply when the other free parameters are fixed as discussed.

This review is too short to cover all possible low mass scenarios for scalars at colliders. We already discussed quite a few new physics extensions including a possible discovery prospect in \cite{Robens:2022zgk}.
Additional work since that review include e.g. updates of studies with invisible final states or $W^+ W^-$ final states \cite{Zarnecki:2025sfy}. Other models that could in principle be of interest include the NMSSM \cite{Cao:2024axg}, a complex $S_3$ symmetric 3HDM \cite{Kuncinas:2023ycz}, as well as a $\mathbb{Z}_2$ symmetric Georgi-Machecek model \cite{deLima:2022yvn}.  In addition, various works try to accomodate a novel scalar with a mass around 95 \GeV that is in agreement with LEP and LHC excesses and the possible discovery potential; see e.g. \cite{Biekotter:2023jld,Sharma:2024vhv,Dong:2025exu,Dong:2025orv,Chen:2025vtg,Chang:2025bjt}. A general overview on state of the art on extended scalar sectors can be found in \cite{Robens:2025nev}.

\begin{center}
  \begin{figure}[htb!]
    \begin{center}
      \includegraphics[width=0.6\textwidth]{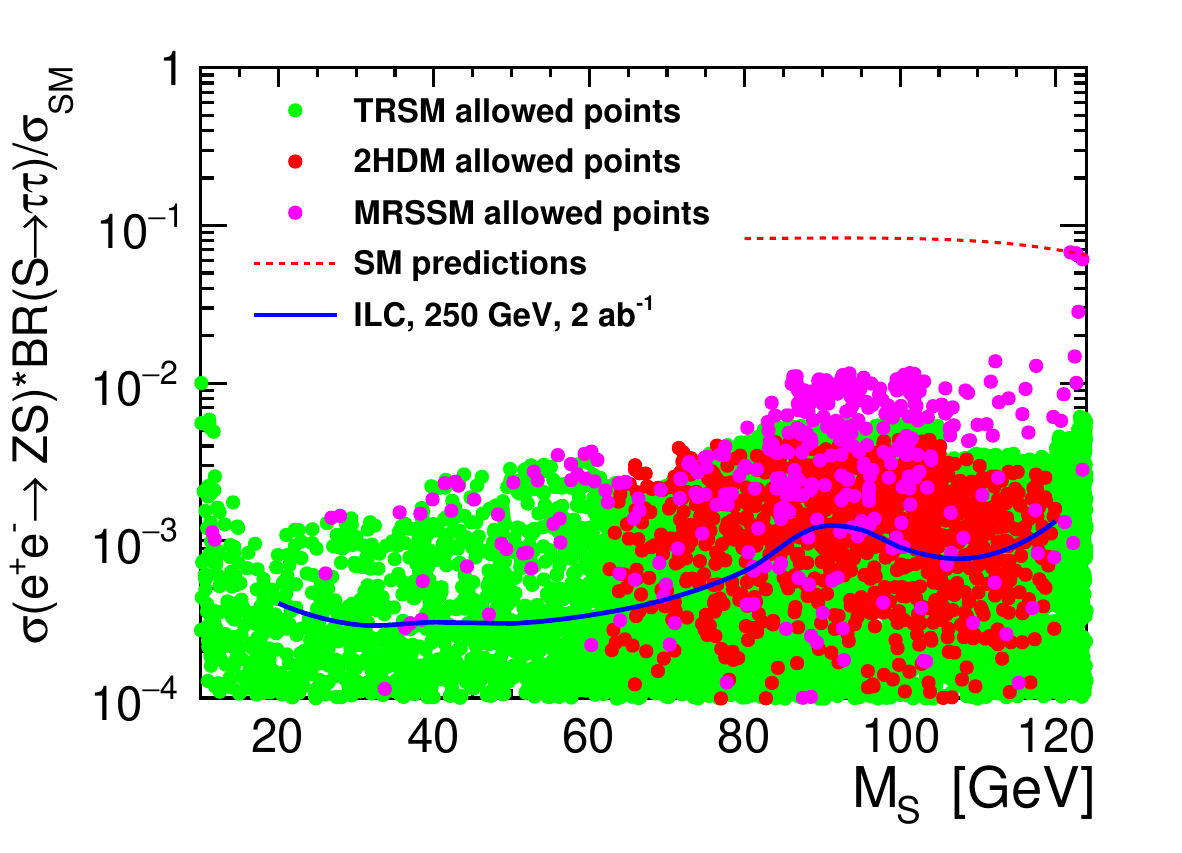}
    \end{center}
    \caption{\label{fig:taucomp} Taken from \cite{Altmann:2025feg} for $\tau^+\,\tau^-$ final states, viability of different models.}
    \end{figure}
  \end{center}

Finally, we show the possible reach of an additional scalar decaying into ditau final states and compare to the parameter space recently still allowed by theoretical and experimental constraints. This is displayed in figure \ref{fig:taucomp}, which compares the reach of the searches discussed previously with currently available parameter space from various new physics scenarios, in particular the TRSM, a 2HDM, as well as the MRSSM which corresponds to the MSSM with a continuous R-symmetry (see e.g. \cite{Diessner:2019bwv,Kalinowski:2024uxe}). It is obvious that given currently allowed regions, this search can severly constrain the parameter space of the model.

\section{Conclusion and outlook}
I briefly discussed the prospects of investigating low mass scalars at $e^+\,e^-$ facilities, with a focus on scalar-strahlung processes as well as the respective decays into various SM final states. Given the current landscape and development of the field, these processes are still underinvestigated and should be given more focus in the upcoming years. For the time being, several studies exist that show promising results for the investigation of scalar strahlung. In addition, there are quite a few well-motivated new physics scenarios that still provide viable candidates for low mass scalars in the regions that are of interest here.

Apart from the processes discussed above, also processes with dark matter final states in extended scalar sectors have been investigated, see e.g. \cite{Kalinowski:2018kdn,Bal:2025nbu} for various collider setups and center-of-mass energies. For these scenarios, such machines also constitute good tools for discovery or alternatively constraining the respective models parameter spaces.

I equally briefly mentioned the important connection to electroweak phase transitions in the investigation of such final states at low center of mass energy lepton colliders. Also for these scenarios such machines proove important tools to discover or exclude regions in such models parameter spaces.

\section{Acknowledgements}
TR acknowledges financial support from the Croatian Science Foundation (HRZZ) project " Beyond the Standard Model discovery and Standard Model precision at LHC Run III", IP-2022-10-2520. This contribution was supported by the National Science Centre, Poland, under the OPUS research project no. 2021/43/B/ST2/01778.
\bibliography{lit,lit_lw,lit_epi}
\end{document}